\documentclass{emulateapj}
\usepackage{float}
\usepackage{epsfig,natbib}
\def\lap{\lower.5ex\hbox{$\; \buildrel < \over \sim \;$}}
\def\gap{\lower.5ex\hbox{$\; \buildrel > \over \sim \;$}}

\begin{document}

\title{The Ages of High Mass X-ray Binaries in NGC~2403 and NGC~300}

\author{Benjamin F. Williams\altaffilmark{1},
Breanna A. Binder\altaffilmark{1},
Julianne J. Dalcanton\altaffilmark{1},
Michael Eracleous\altaffilmark{2},
Andrew Dolphin\altaffilmark{3}
}
\altaffiltext{1}{Department of Astronomy, Box 351580, University of Washington
, Seattle, WA 98195; ben@astro.washington.edu; bbinder@astro.washington.edu;jd@astro.washington.edu}
\altaffiltext{2}{Pennsylvania State University; mce@astro.psu.edu}
\altaffiltext{3}{Raytheon; adolphin@raytheon.com}


\begin{abstract}

We have examined resolved stellar photometry from {\it HST} imaging surrounding 18 high-mass X-ray binary (HMXB) candidates in NGC~300 and NGC~2403 as determined from combined {\it Chandra} /{\it HST} analysis.  We have fit the color-magnitude distribution of the surrounding stars with stellar evolution models.  All but one region in NGC~300 and two in NGC~2403 contain a population with an age between 20 and 70 Myr.  One of the candidates is the ultraluminous X-ray source (ULX) in NGC~2403, which we associate with a 60$\pm$5~Myr old population.  These age distributions provide additional evidence that 16 of these 18 candidates are HMXBs.  Furthermore, our results suggest that the most common HMXB age in these galaxies is 40--55~Myr. This preferred age is similar to observations of HMXBs in the Small Magellanic Cloud, providing new evidence of this formation timescale, but in higher metallicity populations.  We suggest that this preferred HMXB age is the result of the fortuitous combination of two physical effects.  First, this is the age of a population when the greatest rate of core-collapse events should be occurring, maximizing neutron star production.  Second, this is the age when B stars are most likely to be actively losing mass.  We also discuss our results in the context of HMXB feedback in galaxies, confirming HMXBs as a potentially important source of energy for the interstellar medium in low-mass galaxies.

\end{abstract}

\section{Introduction}

High mass X-ray binaries (HMXBs) provide one of the best probes of the
endpoint of massive stars' evolution.  These binaries consist of a
compact object accreting material from a massive companion, allowing
stringent lower limits to be placed on the mass of the compact object
progenitor. Furthermore, the high luminosity of the companion allows
the potential for radial velocity measurements to determine the binary
mass ratio.  In cases where the HMXB is X-ray bright, the X-ray
variability and spectral properties yield information about the nature
of the accretion flow onto the compact object.

Because HMXBs are such powerful laboratories for understanding both
high-mass and binary stellar evolution, the known X-ray bright systems
in the Galaxy have been studied in great detail \citep[e.g.][and
references therein]{tanaka1996,lewin2006,remillard2006}.  The Small
Magellanic Cloud's (SMC's) rich HMXB population \citep{yokogawa2003}
has also been well-studied, providing a probe of HMXBs at low
metallicity \citep{antoniou2009spec,antoniou2009opt,antoniou2010}.

Measurements of formation timescales of HMXBs provide key constraints
for binary evolution models.  Determining the ages of HMXB systems in
the Milky Way is difficult due to distance uncertainties and high
extinction, as most HMXBs are in the Galactic plane.  The most
fruitful studies to date have focused the SMC, which find a typical
formation timescale of 25--60 Myr \citep{antoniou2010}.  This
timescale would explain the SMC's richness in HMXB, since it
experienced a strong episode of star formation $\sim$50~Myr ago
\citep{shtykovskiy2007,antoniou2010}.

It is difficult to determine conclusively why HMXBs would preferably
form on a $\sim$50~Myr timescale. One possible explanation is that
this is the time when B stars -- the most common type of secondary in
HMXBs -- shed mass at the highest rate.  In the Galaxy, B star
activity reaches a maximum between 25 and 80~Myr \citep{mcswain2005},
in broad agreement with the HMXB age distribution in the SMC.

Herein, we turn our attention to two new galaxies, NGC~300 and
NGC~2403, where we may learn about the timescales of HMXB formation at
a metallicity that is higher than the SMC and more characteristic of
the older stars in galactic disks (-0.7$\lap$[Fe/H]$\lap$-0.3,
\citealp{bresolin2009,gogarten2010,garnett1997,barker2012,williams2013}),
which may play a role in early HMXB feedback \citep{justham2012}.
NGC~300 is a SA(s)d type, at a distance of 2.0 Mpc
\citep{dalcanton2009}.  It is nearly face-on
\citep[42$^{\circ}$][]{carignan1985}, and of intermediate metallicity.
Thus the complications of dust extinction are minimized, and it probes
an interesting mass-metallicity regime between the SMC and the Galaxy.
NGC~2403 is a similar type galaxy with a somewhat larger mass,
distance \citep[3.3~Mpc][]{dalcanton2009}, with similar metallicity,
and dust content \citep{williams2013}.

Recently, deep {\it Chandra} observations of NGC~300 and NGC~2403 were
analyzed as part of the Chandra Local Volume (CLV) survey
\citep[][B. Binder et al., in preparation]{binder2012}.  This study
included a new method for identifying strong HMXB candidates using a
combination of the {\it Chandra} data and overlapping {\it HST}
imaging available from the ACS Nearby Galaxy Treasury project
\citep[ANGST][]{dalcanton2009}.  In total, 18 HMXB candidates were
found within the boundaries of deep {\it HST} images of NGC~300 and
NGC~2403. These X-ray sources all have a blue ($V{-}I{<}0$ equivalent)
optical counterpart candidates that fall within the {\it Chandra}
error circle.  The brightest of these candidates were taken to produce
the histogram of optical magnitudes in Figure~\ref{image}. The {\it
HST} data provide deep resolved stellar photometry near the HMXB
positions, which we fit with stellar evolution models to search for
young coeval populations to constrain the HMXBs' ages.

The paper is organized into 4 sections.  Section 2 discusses the data
and analysis techniques used to make our age distribution
measurements.  Section 3 discusses the results of the measurements,
including implications concerning the formation of HMXBs.  Finally,
Section 4 provides a brief summary of our conclusions. We assume
distances of 2.0 Mpc and 3.2 Mpc to NGC~300 and NGC~2403 respectively
throughout the paper, and all X-ray luminosities are quoted for the
0.35-8.0~keV energy range.

\section{Data Reduction and Analysis}

\begin{figure*}
\centerline{\psfig{file=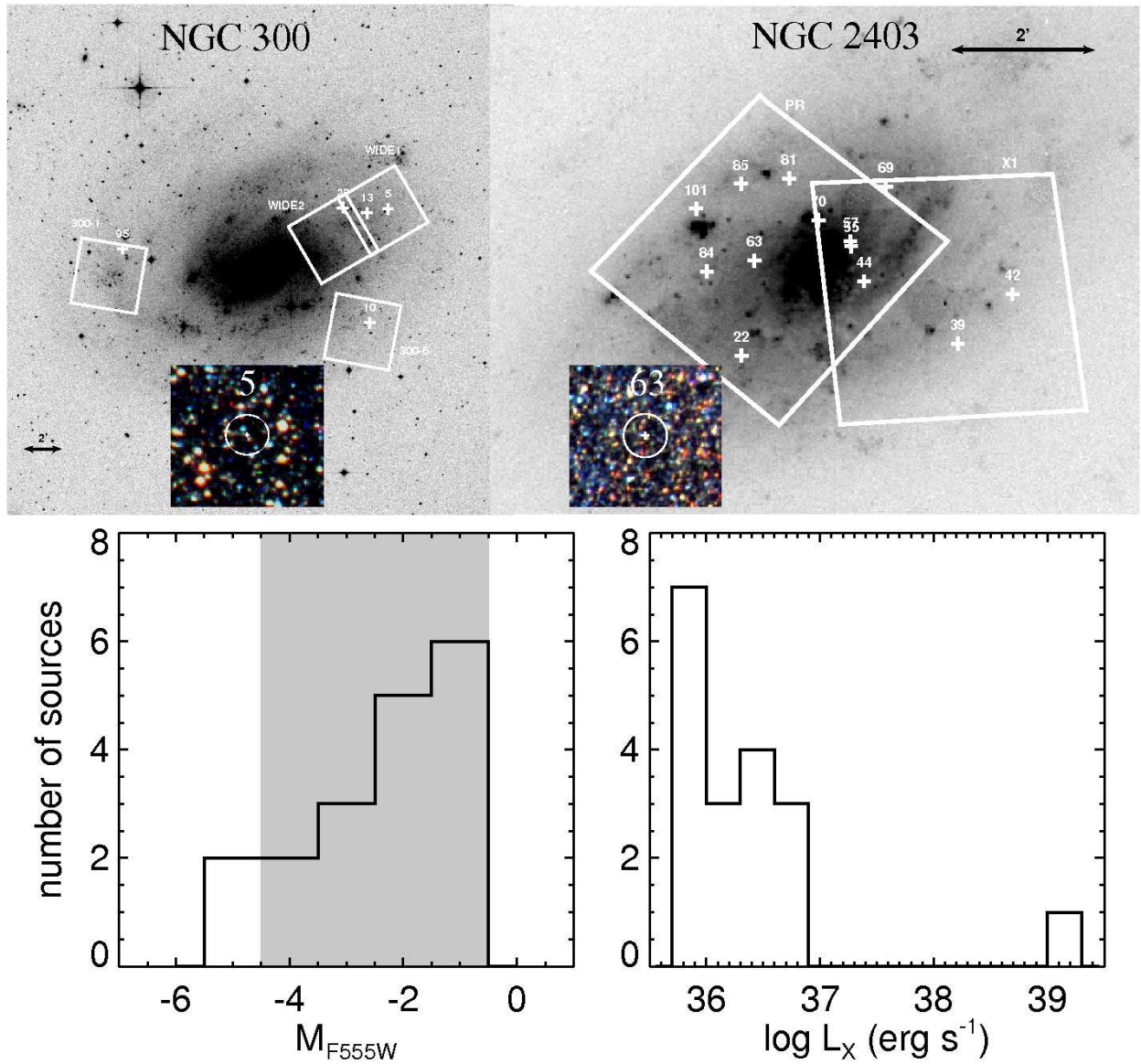,width=6.5in,angle=0}}
\caption{{\it Top:} Footprints of our sample {\it HST} fields shown on
Digitized Sky Survey images of NGC~300 (left) and NGC~2403 (right).
North is up and East is left. Fields are labeled with shortened
versions of their names given in Table~\ref{table}. Crosses mark the
relevant HMXB candidate locations. Inset are 5$''$ sections of the
{\it HST} images around one HMXB candidate from each galaxy, to show
the uniformity of the stellar density on these spatial scales.  {\it
Bottom Left:} Histogram of the absolute F555W magnitudes of the
brightest optical sources coincident with the HMXB candidates in our
sample.  The shaded region shows the typical luminosities of B-type
stars.  {\it Bottom Right:} Histogram of the observed 0.35-8~keV X-ray
luminosities of our sample.  All but the ULX are in the luminosity
range typical of HMXBs in outburst.}
\label{image}
\end{figure*}

There were 4 relevant archival {\it HST} fields in NGC~300 and 2 in
NGC~2403.  Two of these fields were part of the ANGST project
\citep[GO-10915; NGC0300-WIDE1, NGC0300-WIDE2][]{dalcanton2009}, the
other four were taken from the {\it HST} archive (GO-9492 -- NGC300-1,
NGC300-6; GO-10579 -- NGC2403-X1; GO-10182 -- SN-NGC2403-PR).  The
footprints of these fields are shown on Digitized Sky Survey images of
NGC~300 and NGC~2403 in Figure~\ref{image}.  The details of the
filters, exposure times, and depths are given in Table~\ref{table}.

Through cross-correlation of {\it Chandra} and {\it HST} catalogs
\citet{binder2012} found 5 HMXB candidates in NGC~300 and 13 HMXB
candidates in NGC~2403 that fell within {\it HST/ACS} imaging fields.
The distributions of optical magnitudes and X-ray luminosities of
these candidates are shown in Figure~\ref{image}.  The vast majority
of the optical counterpart candidates have optical magnitudes the are
typical of B-type stars, and the X-ray luminosities are typical of
HMXBs in outburst.  The {\it Chandra} data are not sensitive enough to
reach quiescent HMXBs at the distance of NGC~300.  One source is
clearly ultraluminous.  This is the NGC~2403 ULX, which has an optical
counterpart candidate with M$_{F555W}$=-2.6.

\subsection{Photometry}

\begin{figure*}
\centerline{\psfig{file=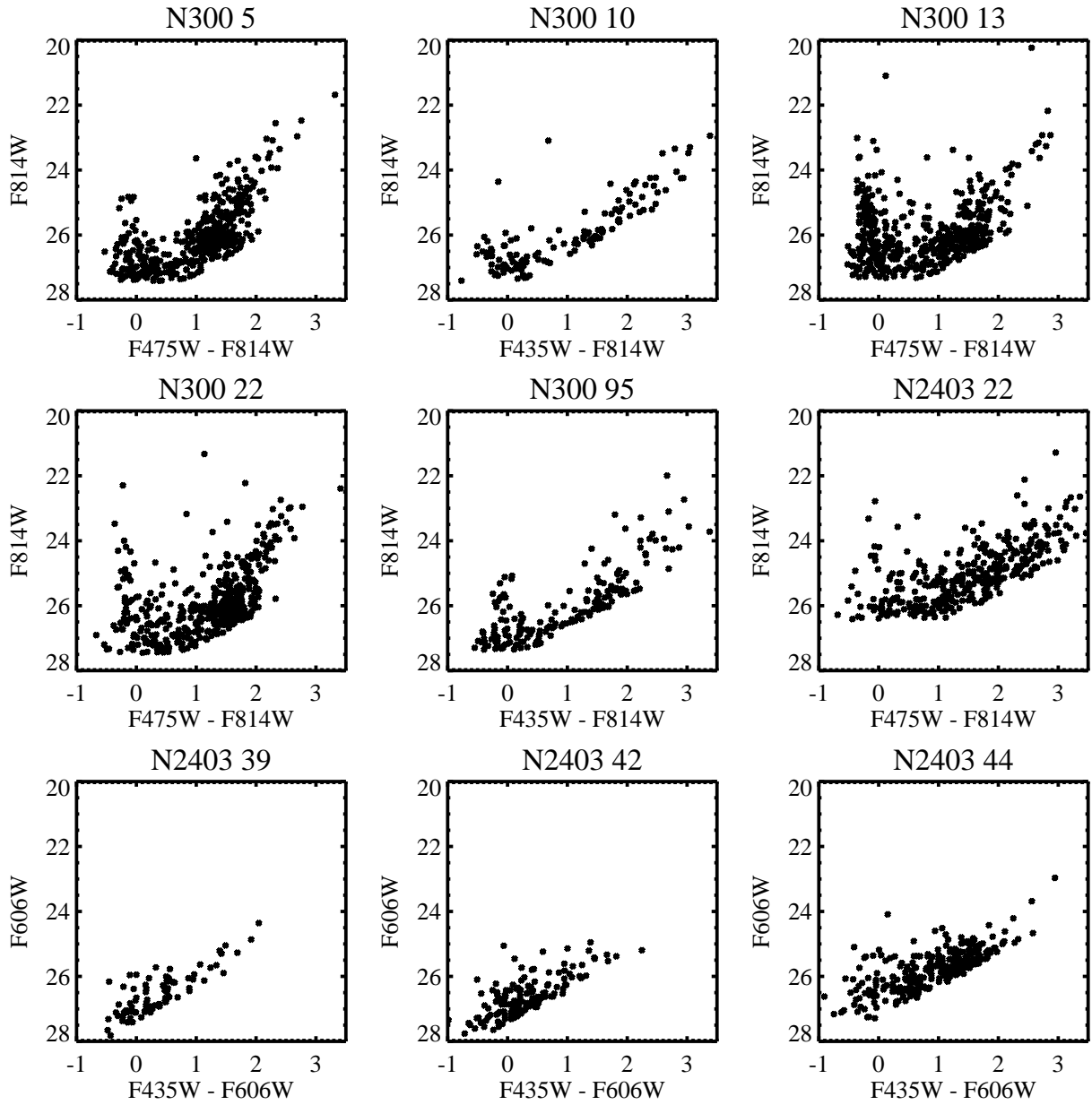,height=6.5in,angle=0}}
\caption{Color-Magnitude diagrams for 9 of the regions in our study.
These represent all of the measured stars within 50~pc of the
locations of 9 of the HMXB candidates marked with crosses in
Figure~\ref{image}.  The diagrams extend fainter than the 50\%
completeness limit of the data, as stars are still measured at those
magnitudes at low levels of completeness, making them unusable for our
model fitting.}
\label{cmds1}
\end{figure*}

\begin{figure*}
\centerline{\psfig{file=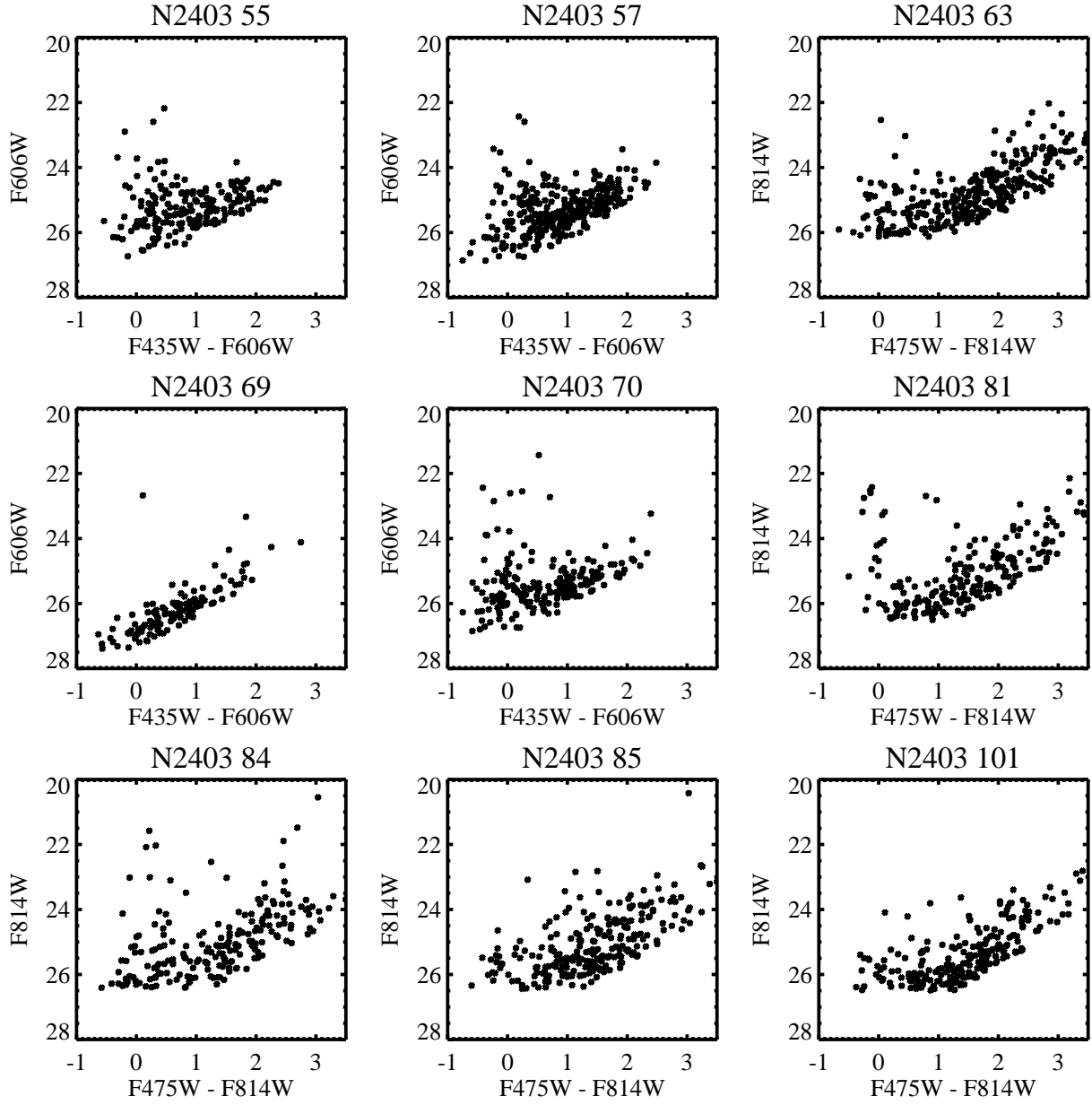,height=6.5in,angle=0}}
\caption{Same as Figure~\ref{cmds1}, but for the remaining 9 of the
regions in our study.}
\label{cmds2}
\end{figure*}

All photometry and artificial star tests were measured and performed
as described by the ANGST project paper \citep{dalcanton2009}.
Briefly, all photometry was performed using DOLPHOT, an updated
version of HSTphot optimized for ACS photometry.  Photometry was
culled based on signal-to-noise ratio and the quality parameters of
sharpness and crowding, as described in \citet{dalcanton2009}.  We
selected only the photometry from a circular region with a radius of
50 pc around the positions of the HMXBs (5.1$''$ and 3.2$''$ in
NGC~300 and NGC~2403, respectively).  The CMDs of these regions are
shown in Figures~\ref{cmds1} and \ref{cmds2}.  Fake star tests were
taken from larger regions (25$''$ and 15$''$) in order to gain a large
number of tests to improve statistics of the photometric uncertainties
and completeness as a function of color and brightness.  The {\it HST}
images of one source in each galaxy are given in Figure~\ref{image},
showing that the stellar density and extinction does not vary strongly
within these relatively small spatial scales.

\subsection{Color-Magnitude Diagram Fitting}

We fit the color-magnitude diagrams for each of the fields in our
study using the software package MATCH \citep{dolphin2002}.  The
overall technique for our fitting, as it has been applied for all
ANGST papers, is described in detail in \citet{williams2009a};
however, there have been some changes in the method for uncertainty
estimation.  Output star formation rates are renormalized to a
\citet{kroupa2001} initial mass function.  Furthermore, we find the
best-fitting mean extinction to each location with the distance fixed
to 2.0~Mpc and 3.2~Mpc for NGC~300 and NGC~2403, respectively
\citep{dalcanton2009}.  We found the best fitting extinction to be
consistent with the foreground value from \citet{schlegel1998}.  In
addition, we attempted to improve the model fits by including
differential reddening in the model Hess diagrams by spreading the
model photometry along the reddening line using the MATCH {\tt dAv}
flag; however, we found no improvement and therefore no evidence for
significant differential reddening in these locations, suggesting
these are not extremely young active regions associated with O-type
stars.  With these distance and extinction values applied, we then run
a series of 100 Monte Carlo (MC) tests.

To assess the full combination of systematic errors due to model
deficiencies as well as random errors due to the depth and size of
the sample, realizations of the best-fitting model solution are fitted
with the models shifted in bolometric magnitude and effective
temperature \citep{dolphin2012}.  These shifts account for the
uncertainties due to any potential systematic offsets between the data
and models.  This total uncertainty was measured for one location to
assess the total uncertainties in the rest of outer portions of the
galaxy.

When searching for differences between different locations in the same
galaxy, systematic uncertainties due to offsets between the models and
the data are not of concern, since these systematics will affect all
fields.  To estimate the {\it relative} uncertainty for differences
between two location analyzed using the same models, we need only to
assess the random errors due to the depth and size of the sample.
Therefore, for the other locations, our MC error analysis did not
include shifts between the models and data.

\section{Results and Discussion}

\subsection{Source 42: The ULX}

One interesting test case is NGC~2403 source 42, which is a
well-studied ULX that is thought to have a black hole primary
\citep{trinchieri1987,swartz2004,feng2005,isobe2009}.  The black hole
mass is estimated from {\it Suzaku} X-ray spectra to be
10--15~M$_{\odot}$, assuming radiation near the Eddington limit
\citep{isobe2009}.  There is no confirmed optical counterpart.  Our
analysis suggests that it resides in a region with no stars younger
than 50~Myr, but with a significant population of stars with ages
60$\pm$5~Myr.  Since the progenitor of the black hole likely had a
main sequence lifetime of $\lap$10~Myr, our result suggests that this
ULX either has had a very long X-ray lifetime compared to the lifetime
of the primary, or the ULX is so bright now because it has a secondary
that is in a state of rapid mass-loss, as in the model of
\citet{king2001}. {\it Spitzer} imaging from the SINGS
\citep{kennicutt2003} program shows no obvious counterpart in the near
or mid infrared, so it does not appear to be an IR-bright asymptotic
giant branch star, shedding its envelope.  In any case, if the ULX is
associated with this 60~Myr population, then the secondary likely has
a zero-age main-sequence mass of $\lap$7 M$_{\odot}$.

\subsection{A Peak in the HMXB Age Distribution}

\begin{figure*}
\centerline{\psfig{file=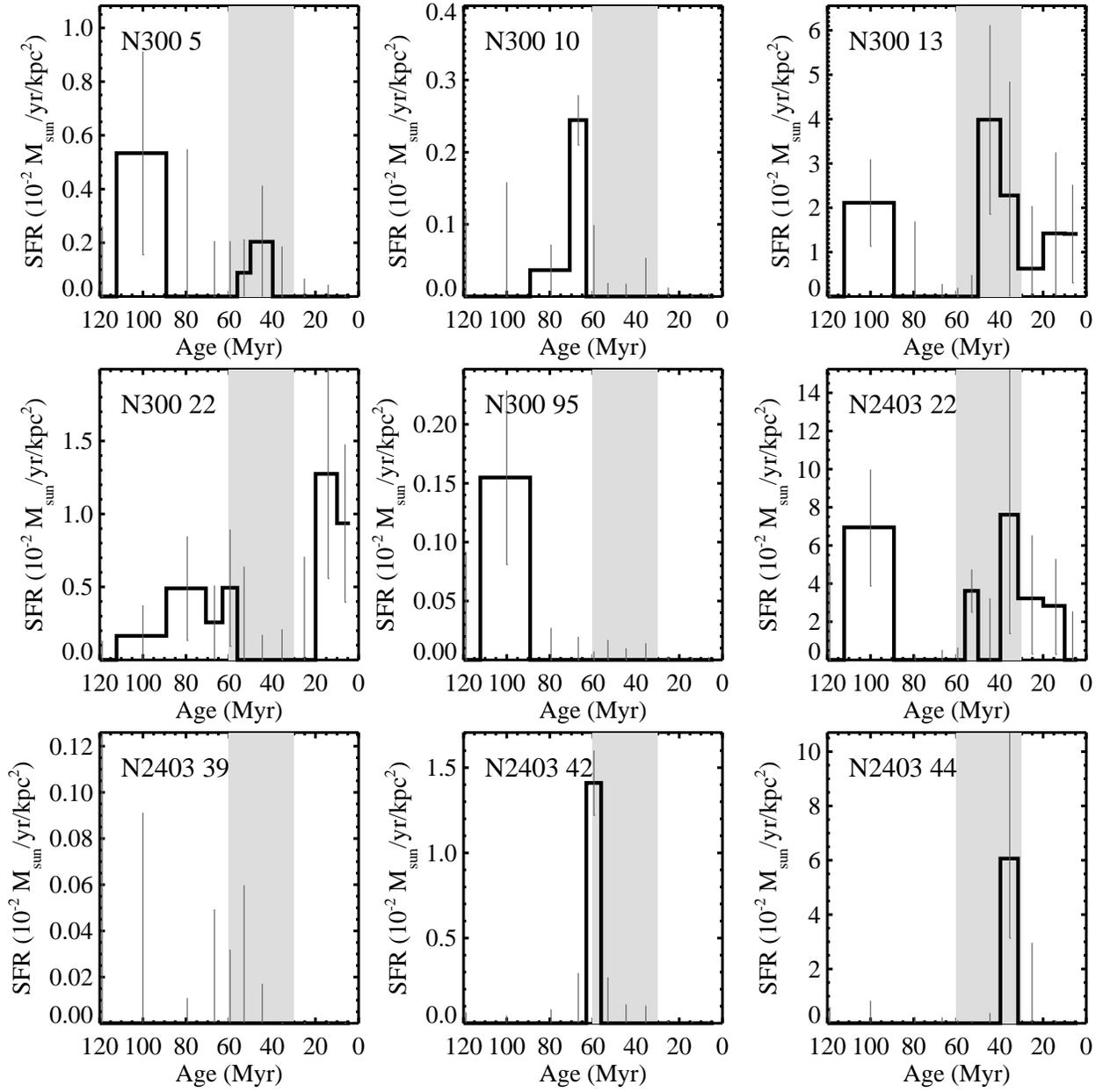,width=6.5in,angle=0}}
\caption{Recent SFHs resulting from model fits to the CMDs shown in
Figure~\ref{cmds1}.}
\label{sfhs1}
\end{figure*}

\begin{figure*}
\centerline{\psfig{file=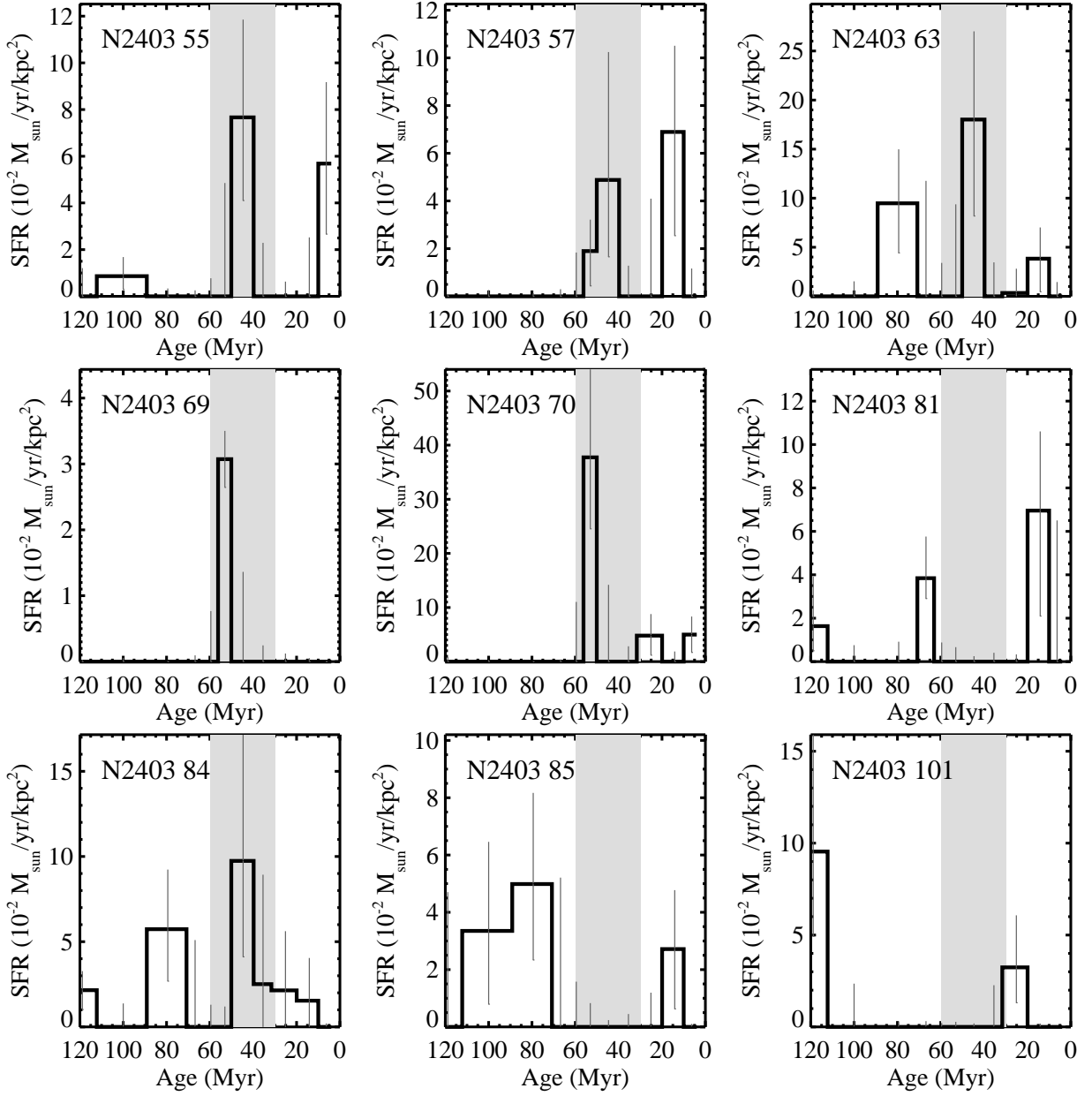,width=6.5in,angle=0}}
\caption{Same as Figure~\ref{sfhs1}, but for the remaining 9 HMXB
candidates shown in Figure~\ref{cmds2}.}
\label{sfhs2}
\end{figure*}

Turning now to the broader sample, in Figures~\ref{sfhs1} and
\ref{sfhs2}, we show the recent star formation histories (SFHs) for
the regions surrounding all 18 HMXB candidates.  For reference, we
highlight the age range between 30 and 60 Myr.  All but 3 (N300-95,
N2403-39, and N2403-85) show a significant peak in between 20 and 70
Myr. One of these (NGC2403-85) shows a population of 10--20~Myr, and
therefore is still a potential HMXB. HMXBs are expected to appear in
populations even younger than 10--20 Myr. The population synthesis
models of \citet{vanbever2000} show that substantial numbers of HMXBs
appear 4-5 Myr after the beginning of star formation. A possible
evolutionary scenario for the formation of such an HMXB on a time
scale of 5-6 Myr is described in \citet[assuming that the binary
survives the first supernova]{belczynski2008}. Observationally, a case
in point is provided by the Galactic cluster Westerlund~1, which has
an age of $\sim$4~Myr. This cluster has several Wolf-Rayet binaries
(potential progenitors of HMXBs), one candidate HMXB, and a magnetar,
indicating that a number of compact objects have already formed from
supernovae \citep[see][]{muno2006,clark2008}. 

N300-95 has a 100~Myr population, pushing the old-age limits for an
HMXB, and N2403-39, while showing no significant population younger
than 100 Myr, has upper limits that allow the possibility of some
young stars in the region.  The X-ray sources in regions with very
little recent star formation could also be LMXBs, since all of these
regions also contain significant old populations. However, if for
example, N2403-39 is an LMXB, then the counterpart is not the blue
star that falls in the Chandra error circle.

\begin{figure*}
\centerline{\psfig{file=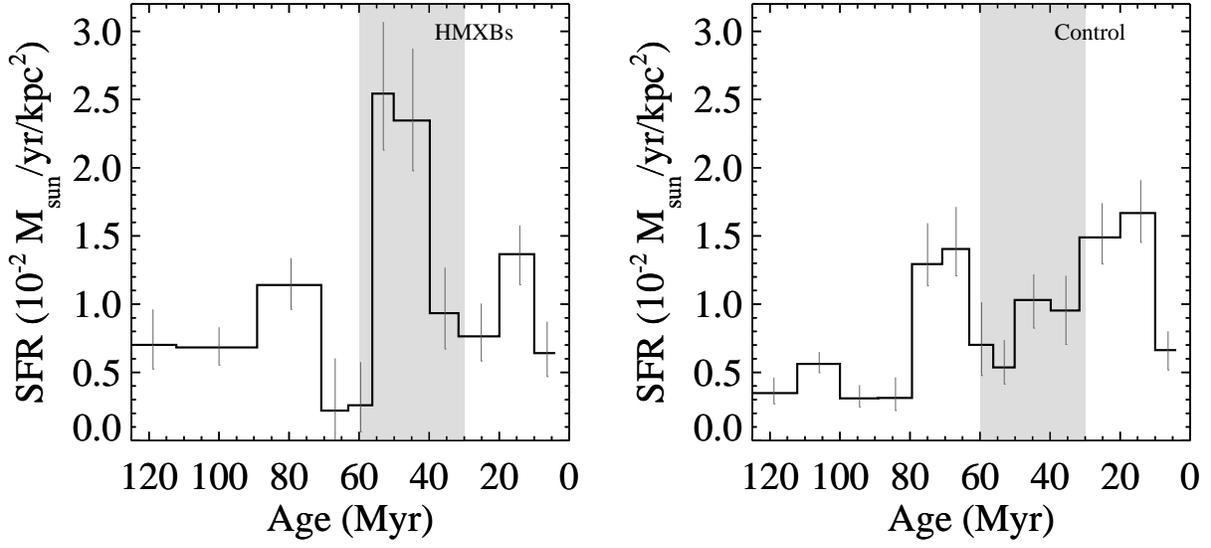,width=6.5in,angle=0}}
\caption{{\it Left}: Mean of all SFHs shown in Figure~\ref{sfhs1} and
\ref{sfhs2}. {\it Right:} Mean of SFHs from 30 random locations in the
{\it HST} fields used for the study.}
\label{sum}
\end{figure*}

When we combine the results from all of the candidates in our sample,
as shown in Figure~\ref{sum}, we find a prominent enhancement in the
age distribution at 40--55 Myr. This result provides more evidence
that 40--55 Myr is a common age for HMXB systems, and therefore may
represent the post star-formation epoch with the largest number of
HMXBs per unit stellar mass formed.  Furthermore, this epoch would
presumably also be the interval of maximum feedback from HMXB jets,
outflows, and high-energy radiation.

To quantify the likelihood of such a peak in a summed age distribution
appearing by chance, we fit photometry from a control sample.  This
control sample consisted of photometry drawn from 18 regions within
0.5$''$ of randomly drawn bright blue stars in the {\it HST} fields.
We required the same number of regions per field as the HMXB sample.
We ran these random locations through our fitting technique and summed
the results, finding no clear peak in the age distribution
(Figure~\ref{sum}).  We then ran this test 1000 times to see if we
ever randomly recovered a peak similar to that seen in the age
distribution of the HMXB sample.  We found that 1.5\% of our trials
resulted in a peak of at least a factor of 2.35 in 2 adjacent bins
between the ages of 20 and 140 Myr (as seen in the HMXB sample).  A
smaller number (0.4\%) of our trials had such peaks as old as the
40--55~Myr peak seen in the HMXB distribution.  Thus, a peak like that
seen in the HMXB sample is a 3$\sigma$ outlier in our trials, roughly
consistent the with uncertainty estimates shown in Figure~\ref{sum}.

\subsection{Implications for Binary Evolution Theory}

The standard model of massive binary evolution suggests that in a
massive binary (e.g. B and O stars or A and O stars), the more massive
star evolves first and transfers mass to the less massive companion
when it overflows its Roche lobe. The core of the massive star
subsequently collapses, forming a neutron star or black hole. If the
binary remains bound after the core-collapse supernova, it may become
an HMXB \citep[e.g.][]{vandenheuvel1973,vandenheuvel1976}.

After mass accretion, the less massive star is spun up due to the
addition of high angular momentum material
\citep{pols1991,vanbever1997}, and subsequent evolution occurs at
higher temperatures \citep[e.g.,][]{vanbever1998}. The rapid rotation
pushes the wind of this hot star into the equatorial plane and leads
to the formation of an equatorial outflow disk \citep{bjorkman1993},
which makes this an active B-star (Be star).  As this mass is
transferred back to, and accreted by, the compact object, X-ray
emission is produced and the binary becomes a Be-HMXB
\citep{vanbever2000}.

Because the B-star is rapidly rotating, the mass outflow is localized
in the equatorial plane.  This ``outflow disk'' increases the mass
transfer efficiency. The resulting X-ray luminosity is low a large
fraction of the time ($<$10$^{35}$ erg s$^{-1}$).  However, these
systems can sometimes reach luminosities of 10$^{37}$ erg~s$^{-1}$
because the outflow disk is unstable and can have episodes of enhanced
outflow \citep[e.g.,][]{reig2007}, which lead to an increase in the accretion
rate (and X-ray luminosity) of the compact object. 

Given this model of HMXB formation, one might expect a correspondence
between HMXB ages and B-star evolution.  Observations suggest that the
fraction of active B stars (Be stars) peaks at 25--80 Myr
\citep{mcswain2005}. Our measurements suggest that the fraction of
HMXBs appears to peak at 40--55~Myr, in good agreement. Thus, the
activity cycle of B-stars appears to be involved in the observed peak
in the age distribution.

The 40--55~Myr timescale is also associated with a possible peak in
the neutron star production rate.  Current theory and observations
suggest that stars with zero age main-sequence masses of
7--8~M$_{\odot}$ are the lowest mass stars to undergo core-collapse to
form neutron stars \citep[e.g.,][]{jennings2012}.  According to the
initial mass function \citep{kroupa2001}, for a given star forming
episode, there are always more lower mass stars than higher mass ones.
Therefore the highest rate of core-collapse likely occurs when the
lifetime of 7--8~M$_{\odot}$ stars has been reached.  This lifetime is
40--55 Myr according to, for example, the Padova stellar evolution
models \citep{marigo2008}.  Therefore, the production rate of neutron
stars is likely involved in the observed peak in the age distribution
as well.  Thus, both the mass-loss by B stars and the number of
potential HMXB systems are at a maximum at $\sim$50~Myr after star
formation, conspiring to form the peak that we see in the HMXB age
distribution.

\subsection{Implications for HMXB feedback}

Recent attempts to include baryonic physics in galaxy formation models
and simulations have found that the treatment of feedback, energy
injection into the interstellar gas, plays an important role in
shaping not only the mass function \citep[e.g.,][]{benson2003} and
star formation histories of galaxies
\citep[e.g.,][]{stinson2007,quillen2008}, but also their dark matter
halo density profiles \citep{governato2010}.  The necessary timescale
and energetics of the feedback required to reproduce observed galaxy
properties are complex, and as a result, the origin of the feedback
remains controversial.  At least part of the necessary feedback
appears to be related to star formation, especially in low mass
galaxies where no active nucleus is present.

We note that the peak we observe in HMXB activity $\sim$50 Myr after
star formation is generally consistent with HMXBs contributing
significantly to the feedback resulting from star formation.  For
example, \citet{justham2012} suggest that the cumulative energy input
from HMXBs can be similar to that from SNe, and they point out that
HMXB feedback may have the appropriate delay time required to produce
episodic star formation histories of low mass galaxies. Our results
suggest that feedback from HMXBs is maximized $\sim$50~Myr after star
formation, roughly consistent with the delay time required by some
models \citep[e.g.,][]{quillen2008}. However, the delay between star
formation and the transfer of SNe energy into the ISM is a similar
length at low SN-rates \citep{dib2006}, therefore both processes
should be considered potentially important to the feedback budget.

Furthermore, \citet{stilp2013} recently discovered a correlation
between the energetics of the neutral interstellar medium (ISM) and
the star formation rate 30--40~Myr earlier.  This result suggests that
the energy associated with star formation is transferred to the
neutral ISM with roughly this delay time.  The consistency of this
timescale with the delay time of HMXB feedback provides further
evidence that feedback from both HMXBs and SNe are potentially
important for regulating star formation in low-mass galaxies.

While the observed delay time between star formation and HMXB activity
makes HMXB feedback potentially important in low-mass galaxies, the
relatively low energies associated with HMXBs, makes them only minor
contributors to feedback in more massive systems. Massive galaxies
contain supermassive black holes.  Central black holes have masses of
$M_{BH}{\sim}1.4\times10^{-3}M_{gal}$ \citep{fabian2012}.  Assuming
the energy released in producing the central black hole is
$E_{BH}=0.1M_{BH}c^2$, putting $E_{BH}$ in units of ergs and $M_{BH}$
in units of solar masses yields
$E_{BH}=2.5\times10^{50}M_{gal}/M_{\odot}$.  We can also determine
$E_{HMXB}$ as a function of $M_{gal}$ by assuming that $E_{HMXB}/t$,
where $t$ has units of seconds, is equivalent to the X-ray luminosity
($L_X$), and setting $SFR_s{\times}t<M_{gal}$, where $SFR_s$ is the
star formation rate in units of $M_{\odot}~s^{-1}$.  We can now apply
the relation between $L_X$ and $SFR$ from \citet{grimm2003}, which is
$L_X=6.7\times10^{39}SFR$, where $L_X$ is in units of erg~s$^{-1}$ and
$SFR$ is in units of $M_{\odot}~yr^{-1}$.  Substituting $SFR_s$ for
$SFR$ yields $L_X=2.1\times10^{47}SFR_s$; then applying
$M_{gal}/t>SFR_s$ and $E_{HMXB}/t=L_X$ gives
$E_{HMXB}<2.1\times10^{47}M_{gal}/M_{\odot}$. Thus,
$E_{BH}/E_{HMXB}{\gap}10^{3}$, making $E_{HMXB}$ a very minor
contributor to the global evolution of massive galaxies.

\section{Conclusions}

We have performed model fits to the color-magnitude distribution of
the stars within 50~pc of HMXB candidates in NGC~300 and NGC~2403 that
have adequate {\it HST} archival data.  The ULX in NGC~2403 is one of
these sources, and it appears to be associated with a population of
60$\pm$5 Myr in age.  The total of the resulting recent star formation
histories show a significant peak between 40 and 55 Myr.  This peak is
coincident with the timescale for HMXB formation as seen in the SMC.
Using these results, we can infer that N300-5, N300-13, N300-22,
N2403-22, N2403-42, N2403-44, N2403-55, N2403-57, N2403-63, N2403-69,
N2403-70, and N2403-84 are the strongest HMXB candidates, and that the
timescale for maximum HMXB activity is similar to that seen in the
SMC.

The formation of an HMXB requires the production of a neutron
star or black hole through a core-collapse supernova of the primary
and mass-loss by the secondary to fuel the accretion process.  Due to
a fortuitous coincidence, both the age of a population when B-star
mass loss is at a maximum \citep[e.g.,][]{mcswain2005} and the age
when neutron star production is a maximum
\citep[e.g.,][]{jennings2012} are $\sim$50~Myr.  Thus, the peak we
observe in the HMXB age distribution strengthens recent observations
and theory of the evolution of massive stars in binary systems.

Finally, this timescale for the peak of HMXB activity after the onset
of star formation may help to address some open questions in the
interplay between star formation and galaxy evolution.  First, it is
consistent with the delay between star formation and feedback
processes required by some models to reproduce episodic star formation
in low-mass galaxies.  Second, it is consistent with the delay time
between star formation and delivery of energy to the ISM observed in
nearby low-mass galaxies.

Support for this work was provided through Chandra Award Number
AR2-13005X issued by the Chandra X-Ray Observatory Center, which is
operated by the Smithsonian Astrophysical Observatory for and on
behalf of the National Aeronautics Space Administration under contract
NAS8-03060.  Additional funding was provided by grants GO-10915 and
GO-11986 from the Space Telescope Science Institute, which is operated
by the Association of Universities for Research in Astronomy,
Incorporated, under NASA contract NAS5-26555.


\clearpage

\begin{deluxetable}{ccccccc}
\tablewidth{14.5cm}
\tablecaption{Summary of Data and Photometry Measurements}
\tabletypesize{\footnotesize}
\tablehead{
\colhead{Proposal} &
\colhead{Target} &
\colhead{Camera} &
\colhead{Filter} &
\colhead{Exposure (s)}&
\colhead{Stars} &
\colhead{$m_{50\%}$} 
}
\startdata
9492 & NGC300-1 & ACS & F435W &   1080 & 94793 &  27.51\\
9492 & NGC300-1 & ACS & F555W &   1080 & 126857 &  27.45\\
9492 & NGC300-1 & ACS & F814W &   1440 & 126857 &  27.10\\
9492 & NGC300-6 & ACS & F435W &   1080 & 72509 &  27.46\\
9492 & NGC300-6 & ACS & F555W &   1080 & 111850 &  27.38\\
9492 & NGC300-6 & ACS & F814W &   1440 & 111850 &  27.07\\
10915 & NGC0300-WIDE1 & ACS & F475W &   1488 & 201775 &  27.84\\
10915 & NGC0300-WIDE1 & ACS & F606W &   1515 & 224152 &  27.84\\
10915 & NGC0300-WIDE1 & ACS & F814W &   1542 & 224152 &  27.04\\
10915 & NGC0300-WIDE2 & ACS & F475W &   1488 & 314579 &  27.29\\
10915 & NGC0300-WIDE2 & ACS & F606W &   1515 & 363837 &  27.04\\
10915 & NGC0300-WIDE2 & ACS & F814W &   1542 & 363837 &  26.53\\
10182 & SN-NGC2403-PR & ACS & F475W &   1200 & 316973 &  26.41\\
10182 & SN-NGC2403-PR & ACS & F814W &    700 & 316973  &  25.47\\
10579 & NGC2403-X1 & ACS & F435W &   1248 & 154848 &  26.84\\
10579 & NGC2403-X1 & ACS & F606W &   1248 & 154848 &  26.32
\enddata
\label{table}
\end{deluxetable}

\end{document}